\documentclass[sigconf, anonymous=false]{acmart}

\usepackage{svg}
\usepackage{soul}
\usepackage{subcaption} %to have subfigures available
\usepackage{makecell}
\AtBeginDocument{%
  \providecommand\BibTeX{{%
    \normalfont B\kern-0.5em{\scshape i\kern-0.25em b}\kern-0.8em\TeX}}}

\setcopyright{acmcopyright}
\copyrightyear{2023}
\acmYear{2023}
\acmDOI{XXXXXXX.XXXXXXX}

\acmPrice{15.00}
\acmISBN{978-1-4503-XXXX-X/18/06}

\begin{document}

%\title{MATH-PHOC: Exploring Deeper Characteristics of Pyramidal Histograms of Characters In Relation To Mathematical Information Retrieval}

% RZ: Proposed retitle, only a suggestion
\title{A Study of PHOC Spatial Region Configurations for\\ Math Formula Retrieval}
%\title{Exploring the Effectiveness and Efficiency of\\PHOC Spatial Region Configurations for Math Formula Retrieval}

%\title{%\rz{Suggested Title Change:}\\
%Math Formula Search using IDF Based Variants of Pyramidal Histograms of Characters}

\author{Matt Langsenkamp}
\authornote{The first two authors contributed equally to this research.}
\email{ml2513@rit.edu}
\affiliation{%
  \institution{Rochester Institute of Technology}
  \streetaddress{1 Lomb Memorial Dr}
  \city{Rochester}
  \state{New York}
  \country{USA}
  \postcode{14526}
}

\author{Bryan Amador}
\authornotemark[1]
\affiliation{%
\institution{Rochester Institute of Technology}
  \streetaddress{1 Lomb Memorial Dr}
  \city{Rochester}
  \state{New York}
  \country{USA}
  \postcode{14526}
}
\email{ma2339@rit.edu}

\author{Richard Zanibbi}
\affiliation{%
\institution{Rochester Institute of Technology}
  \streetaddress{1 Lomb Memorial Dr}
  \city{Rochester}
  \state{New York}
  \country{USA}
  \postcode{14526}
}
\email{rxzvcs@rit.edu }

\renewcommand{\shortauthors}{Langsenkamp, et al.}

\begin{abstract}

A Pyramidal Histogram Of Characters (PHOC) represents the spatial location of symbols as binary vectors. The vectors are composed of 
% RZ ++ next line
levels that  
%of which split 
split a formula into 
% RZ: ++
%a larger number of 
equal-sized regions of one or more types (e.g., rectangles or ellipses). 
% RZ: ++, adapted from: Due to the pyramidal nature of the PHOC model, each subsequent level encodes more specific spatial information than the previous one
For each region type, this produces a pyramid of overlapping regions, where the first level contains the entire formula, and the final level the finest-grained regions.
% RZ: adapted from: Due to the pyramidal nature of the PHOC model, each subsequent level encodes more specific spatial information than the previous one. 
%progressively encode more fine-grain information, by splitting 
%the level 
%a formula into equal-sized regions of one or more types 
%(e.g., horizontal rectangles or concentric ellipses). 
%(e.g., rectangles or ellipses).
% RZ: Removing to keep focus on one task (retrieval, what ARQMath addresses)
%, and has been applied to math formula auto-completion. 
%Due to the pyramidal nature of the PHOC model, each subsequent level encodes more specific spatial information than the previous one. 
In this work, we introduce concentric rectangles for regions, and analyze whether  subsequent PHOC levels encode redundant information by omitting levels from PHOC configurations.
%and comparing their performance to the original configuration as well as 
As a baseline, we include a bag of words PHOC containing only the first whole-formula level.   Finally, using the ARQMath-3 formula retrieval benchmark, we demonstrate that some %amount of 
levels encoded in the original PHOC configurations 
%is 
are redundant, that
% MOVED: implementation and suggests future work, put at end
%Our implementation uses an inverted index over symbols with 64 bit PHOC vectors stored in postings.
PHOC models with rectangular regions outperform earlier PHOC models, and that despite their simplicity, PHOC models are surprisingly competitive with the state-of-the-art.
%Despite its simple spatial representation, a cosine similarity over PHOC vectors has proven competitive with state-of-the-art math formula retrieval models.
PHOC is not math-specific, and might be used for chemical diagrams, charts, or other graphics. 

\end{abstract}


\newcommand{\bma}[1]{{\color{blue} (BMA: {#1})}}
\newcommand{\rz}[1]{{\color{red} (RZ: {#1})}}
\newcommand{\ml}[1]{{\color{orange} (ML: {#1})}}


\received{20 February 2007}
\received[revised]{12 March 2009}
\received[accepted]{5 June 2009}

\maketitle

\section{Introduction}

\par
In Mathematical Information Retrieval (MIR) \cite{survey1, survey2}, search is performed using both mathematical notation and text.
  In documents containing math, mathematical notation is most commonly represented using \LaTeX\space or an XML encoding (e.g., Presentation MathML). In this work we are concerned with queries consisting of a single formula. %search where the only information a user has at their disposal for expressing a query is a single formula; 
  For example, when a user wants to learn or refind the name and definition of a formula (e.g., the Golden Ratio, $\frac{1+\sqrt{5}}{2}$). Our focus is this query-by-expression task.

\par We explore visual retrieval using only symbol labels and locations. This has the advantages of being close to how one might consider formulas when searching based on appearance without deeply considering their structure, and might be closer to how a non-expert may interpret formulas (e.g., where operations and their interaction is unknown). With a symbol-based visual model, formula retrieval may be performed directly using symbols from detected formula regions in a PDF, without the need to interpret formula structure  \cite{avenoso_spatial_2021}. 

\begin{figure}[!tb]
    \centering
    \includegraphics[width=.40\textwidth]{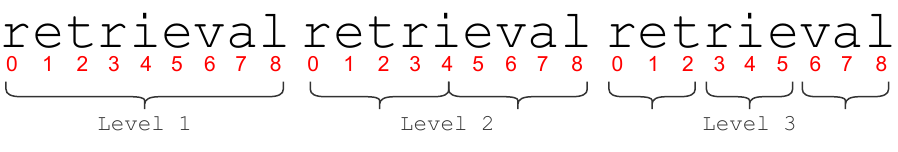}
    \caption{Pyramidal Histogram of Character (PHOC) Regions for the Word {\it retrieval.} Histograms or binary vectors are computed for regions obtained by evenly splitting the character sequence iteratively.
    %e.g., symbol set \{r,e,t,i,v,a,l\} is seen at Level 1, and the two symbol sets \{\{r,e,t,i\},\{i,e,v,a,l\}\} at Level 2. 
    The binary PHOC for `r' in regions above (left-to-right, levels separated by spaces) is $[ 1~10~110 ]$}
    %character of the string below it, and a bracket indicating %what PHOC region it belongs to at each level }
    \label{fig:stringphoc}
\end{figure}

\par The Pyramidal Histogram Of Characters (PHOC) was introduced by Almaz{\'{a}}n et al. \cite{zero_phoc, first_phoc} for \emph{word spotting} in images of handwritten text and text in natural scenes. The original motivation for PHOC was to retrieve words in the presence of OCR errors, by assigning characters to hierarchically-tiled spatial regions (see Figure \ref{fig:stringphoc}). %. Avenoso modifed the representation so that PHOCs where expanded 
Avenoso et al. \cite{avenoso_xy-phoc_2021}  modified PHOC to include vertical and horizontal directions for use in formula retrieval, obtaining symbol locations from Scalable Vector Graphics (SVG) files produced from \LaTeX~using MathJax\footnote{https://www.mathjax.org/}. Despite its simplicity, XY-PHOC scored within 8\% of the top system and exceeding the baseline system by 1.4\% for P$'$@10 in the ARQMath 2021 Formula Retrieval task \cite{ARQ2021}.  The XY-PHOC model was later extended to include elliptical regions, and incorporate a simple form of inverse document frequency (IDF) in scoring \cite{DBLP:conf/clef/LangsenkampMZ22, DBLP:conf/clef/MansouriNAOZ22, avenoso_xy-phoc_2021}. 

% Don't emphasize this here.
%A number of technical and systems errors prevented the new system from achieving the same performance as the original XY-PHOC system.

In this work we adapt this PHOC model through the addition of concentric rectangle regions, and explore the effect of omitting region levels in PHOC vectors. 
%The motivation of skipping levels is 
We believe that some information in adjacent levels is redundant for formula retrieval. Also, omitting bits from levels allows 
%also \rz{allow retrieval to be} more efficient as 
less data to be stored in the inverted index used to retrieve PHOCs for symbols.
%, and less computations need to be performed to score documents. 
The main research questions we seek to answer in this work are as follows.
\begin{enumerate}
    \item Is the addition of concentric rectangle regions beneficial for formula retrieval? 
    \item Does omitting even levels (i.e., levels 2,4,\ldots) from PHOC vectors degrade retrieval effectiveness?
    \item Does omitting all PHOC levels except for final levels with the largest number of regions impact retrieval effectiveness? 
\end{enumerate}

\section{Related Work}
\label{sec:related}

\textbf{Math Formula Retrieval Models}: The earliest formula retrieval systems indexed \LaTeX\space tokens, substituting certain \LaTeX\space structures with special tokens, and then performing retrieval with  traditional textual vector space models such as TF-IDF \cite{DBLP:conf/drr/ZanibbiY11, DBLP:journals/amai/MillerY03, tompaBM25, art} . With this approach, variables and operators are indexed as if they were words in natural language. 
\par Other approaches exploit the inherent visual and semantic graph structures of mathematics. Math can be modeled by graphs including operator trees (OPT), symbol layout trees (SLT), baseline structure trees (BST), and line-of-sight graphs (LOS) \cite{tangentS, Tangent, kennythesis, bst, approach0, kristianto2016mcat}. Approaches such as Tangent\cite{Tangent} index formulas using pairs of SLT symbols with their distances or spatial relationships in the SLT. This technique was later extended to OPT and LOS graphs in Tangent-S and Tangent-V respectively \cite{tangentS, kennythesis}. These graphs all describe relationships between symbols, and whether represent variables or operators, and by indexing symbol on can effectively retrieve using local context within formulas. Candidate formulas are retrieved based on  symbol pairs, and custom sub-tree matching algorithms are used to re-rank the results. 

\par Neural embedding models haven been applied to retrieve formulas \cite{tangentCFT, colbertWEI, mathbert, symbol2vec}. Tangent-CFT \cite{tangentCFT} embeds tuples of symbols in the formulas using their paths in SLT and OPT trees. Systems like MathBERT \cite{mathbert} and MathAMR \cite{mathamr} apply embeddings that also take context into consideration (e.g., surrounding text). 

\par Formula retrieval models that directly use graph structures  and neural embeddings have appealing aspects: graphs-based models utilize structure explicitly, and 
%an intuitive data structure and 
machine learning techniques automatically discover patterns that humans struggle to enumerate. However, both approaches also have drawbacks. When graph structures are used, additional functions are needed to score subgraph matches %which understand the semantics of 
%the graph 
that can be difficult to interpret and design effectively. Embedding models can be expensive and time consuming to train, especially when using modern transformer architectures \cite{parrot}. This does not mean we should abandon graph-based or embedding-based models, but does strongly motivate continuing to explore new, simpler ways to represent formulas for search.

\par\textbf{Pyramidal Histogram of Characters (PHOC)}: Pyramidal Histograms of Characters is a binary representation for character locations in words, first introduced to address the problem of word-spotting in handwritten document images \cite{first_phoc}. A word image is split into horizontal regions of equal size, iteratively increasing the number of regions at each level of splitting. In Figure \ref{fig:stringphoc}, this starts with the whole word (Level 1), then splitting into two halves (Level 2), then three horizontal regions (Level 3) and so on, which produces a pyramid of spatial regions within the word image.  For each character in the word, its PHOC is a binary vector representing the regions containing that character.
%the string at increasing levels and counting the amount of times a certain character appeared in a certain split.
%While a character can appear in a split more than once, it is only counted once if there are any amount present, thus PHOC vectors are binary. 
%Figure \ref{fig:stringphoc} provides an example of the string ``retrieval" split to three levels. 
The model is simple, requiring only symbol labels and their ordering or spatial locations. Spatial locations for characters in images may be defined using connected component analysis in images \cite{second_phoc}, or symbol bounding boxes in vector graphics formats such as PDF \cite{sscraper} or SVG \cite{avenoso_xy-phoc_2021}.

\par PHOC was expanded to two dimensions by Avenoso for use in  formula retrieval and auto-completion in the XY-PHOC model used for the ARQMath-2 competition \cite{avenoso_xy-phoc_2021}. Avenoso added splits perpendicular to the Y-axis (Y direction) and used the horizontal span of characters centered at their centroid to determine region membership in SVG formula images. Splits were expanded to level 5 in both the X and Y direction.  In this model, a formula PHOC vector $\textbf{a}$ concatenates PHOCs for all individual symbols $s$ in symbol vocabulary $V$ in a fixed order. 

Avenoso also modifies the cosine similarity function as shown in Equation \ref{eq:cos} to use bit operations, and take advantage of the query vector being fixed for all candidates. 
%will be the same in all cases and does not need to be recomputed. 
%\par 
In  Equation \ref{eq:cos} $\textbf{a}$ and $\textbf{b}$ are the formula query and formula candidate PHOC vectors respectively, while $| \cdot |_1$ represents the Hamming weight and $\land$ is logical AND. 
\begin{equation}
\label{eq:cos}
\mathit{cos}{(\mathbf{a},\mathbf{b})}\overset{rank}{=} \mathit{bcos}{(\mathbf{a},\mathbf{b})}=\frac{1}{\sqrt{|\mathbf{b}|_{1}}}|\mathbf{a}\land \mathbf{b}|_{1}
\end{equation}
\par For the ARQMath 3 task, this work was extended to allow for new concentric ellipse regions, and the inclusion of inverse document frequencies for symbols. Ellipse regions were added to account for cases where the placement of symbols is inverted, for example $x+y+z=1$ and $1=z+y+x$. 

\begin{figure}
    \centering
    \begin{subfigure}[t]{0.20\textwidth}
        \raisebox{-\height}{\includegraphics[width = 80pt]{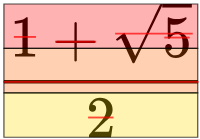}}
        \caption*{Y-Axis Splitting (Level 3)}
    \end{subfigure}
    \hfill
        \begin{subfigure}[t]{0.20\textwidth}
        \raisebox{-\height}{\includegraphics[width = 80pt]{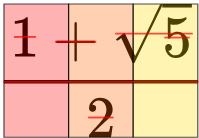}}
        \caption*{X-Axis Splitting (Level 3)}
    \end{subfigure}
    ~\\~\\
    \begin{subfigure}[t]{0.20\textwidth}
        \raisebox{-\height}{\includegraphics[width = 80pt]{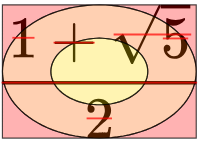}}
        \caption*{Concentric Ellipses (Level 3)}
    \end{subfigure} 
    \hfill
    \begin{subfigure}[t]{0.20\textwidth}
        \raisebox{-\height}{\includegraphics[width = 80pt]{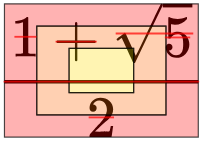}}
        \caption*{Conc. Rectangles  (Level 3)}
    \end{subfigure}

    \caption{PHOC Spatial Regions for the Golden Ratio. Shown is Level 3 for concentric ellipses, X and Y-Axis partitions, and concentric rectangles.}
    \label{fig:golden}

\end{figure}

Figure \ref{fig:golden} shows a formula %$\frac{1 + \sqrt{5}}{2}$ at level 3 split using: 
split up to level 3 using horizontal and vertical splits, nested ellipses, and the concentric rectangles introduced in this paper as a new PHOC region type.
%There was some evidence that this method helped retrieve inverted formulas, but information in the vector still seems dominated by the information in the X and Y directions.
%\par
To represent PHOC configurations, we use the notation of Langsenkamp et al. \cite{DBLP:conf/clef/LangsenkampMZ22} where letters denote region type (x, y, r or o) and integers after letters  indicate the level that the preceding sequence of region types (letters) are expanded to. For example, xy5 denotes a PHOC configuration in which vertically split regions and horizontally split regions are expanded to the fifth level. As another example, x2r7 vertically splits regions to the second level, and rectangular regions are expanded to the seventh level.

\par \textbf{ARQMath Formula Retrieval Task}: In 2020 the ARQMath competition was introduced to further the field of MIR \cite{arq2020}. The competition made use of data from Math Stack Exchange (MSE)\footnote{https://math.stackexchange.com/}, a popular mathematics question answering forum from 2010-2018 as the test collection. Queries where then taken from questions appearing on MSE in 2019, ensuring that no test questions appeared in the training set. Two tasks where established for ARQMath 1; an answer retrieval task and a formula retrieval task. 
%For the answer retrieval task, given a 2019 question from MSE participants had to return the top 1000 answer posts from prior years. 
The formula retrieval task asks participants to return the top 1000 formulas given a formula query taken from a test question post. Returned formulas from all participants where then pooled and evaluated for relevance by math undergraduate and graduate students, taking the context of the post containing a retrieved formula into account. The full ARQMath dataset contains around 28 million formulae; approximately 9.3 million formulas are visually unique. The ARQMath competition was held in 2020, 2021, and 2022 denoted by AQRMath-1, ARQMath-2, and ARQMath-3, respectively; each year taking a new set of test queries from the proceeding calendar year.

\section{Rectangular Regions and\\Level Skipping}
\label{sec:methods}

\par \textbf{Rectangular Regions}: Originally horizontal and vertical regions (see Figure \ref{fig:golden}) were used for formula PHOCs.  Concentric elliptical regions were later added to capture symmetry in symbol positions \cite{DBLP:conf/clef/LangsenkampMZ22} (e.g., so that `x+y' and `y+x' can have similar/identical PHOCs).
In this paper we introduce concentric rectangles: the nested rectangular regions capture similar information as the ellipses, and are less expensive to compute (roughly 50\% faster for rectangles vs. ellipses). Figure \ref{fig:golden} shows the difference between concentric rectangular regions and concentric elliptical regions. 

\par
\begin{figure}
    \centering
    \begin{subfigure}[t]{0.5\textwidth}
        \raisebox{-\height}{\includegraphics[width = 230pt]{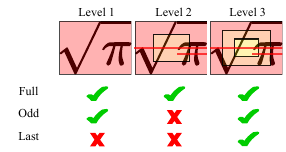}}
        %\caption*{Y-Axis Splitting (Level 3)}
    \end{subfigure}

    \caption{Concentric Rectangular 3-Level PHOC (R3) PHOC with levels shown left-to-right. Check marks indicate  levels present in R3-full, R3-odd, and R3-last PHOC models.}
    %, the R3-odd configuration, and the R3-last configuration.}
    \label{fig:skip}

\end{figure}

 \textbf{Level Skipping}: The pyramidal nature of PHOCs leads to subsequent levels often encoding similar spatial information to the preceedings level. To test whether or not all levels are needed, we produce configurations in which even levels are omitted and configurations in which all levels except for the very last one are omitted. These configurations are named \textit{odd} and \textit{last} configurations respectively. Figure \ref{fig:skip} shows an R3 configuration, with green check-marks indicating the levels present in each of the R3-full, R3-odd and R3-last PHOC model. Note that the R3-full has 6 bits in total, while R3-odd has 4 bits, and R3-last has 3 bits. 

\section{Experimental Setup and  Results}
% \rz{Explain and cite Sakai for definition of the prime metrics -- not all readers know what these are.}
To test whether some PHOC levels are redundant, we perform two experiments. First, we test for %statistically significant results 
differences between various PHOC configurations and X1, which is the same as a Bag of Words (BoW), using precision@k metrics $P'@10$, $P'@5$, and $P'@1$, as well as $MAP'$ and $nDCG'@1000$. Following ARQMath \cite{ARQ3}, we use the "prime" metrics introduced by Sakai \cite{sakai}, that considers only judged documents for scoring.
%to match the conventions used during the ARQMath competition \cite{ARQ3}. 
Next, we test for differences %statistically significant results
when PHOC levels are omitted for `odd' and `last' PHOC configurations (see Figure \ref{fig:skip}) using the same metrics.
%prime metrics $nDCG'@1000$, $MAP'$, $P'@10$, $P'@5$ and $P'@1$. 

\par \textbf{Configuration Selection}: The space of possible PHOC configurations is large. Considering our 4 different region types (X, Y, O and R) and pruning configurations with more than 64 bits (a word size on many modern machines), we find ourselves with 5620 possible configurations. Running indexing and retrieval with the full ARQMath \cite{ARQ3} test collection takes around 5 hours on our systems, and so we cannot test all of these configurations. The purpose of this work is not to find the \textit{best} possible PHOC configuration, but rather to analyze the information signals encoded within certain \textit{levels}. Nonetheless, we need to choose some promising configurations for our analysis. To do this, we restrict the number of regions we will explore, as well as the amount of data indexed when selecting models, and perform a grid search to find three suitable configurations for study.

\par In our experiments we use PHOC configurations yr7, yr7o3 and x5y3r9. These were selected using a grid search over all combinations of X, Y, R, and O regions using at most 9 levels for each region type, and only odd levels for the maximum level. All configurations must also fit within 64 bits. In total there was 272 configurations. Each configuration was used to index the collection of evaluated formulas from the ARQMath \cite{ARQ3} collection, and evaluated using the 2021 ARQMath Task 2 Topics. We then chose the configurations which had the best $nDCG'@1000$ (71.86\%, y7r7) and $P'@10$ (48.97\%, x5y3r9), the official ARQMath rank metrics, and wanted both deep and shallow rank metrics. Finally we took the mean-reciprocal rank of the system rankings for the two metrics, to identify configurations performing well for deep and shallow metrics: x5y3r9 and y7r7 were strongest here again. For additional information, we included the third-best model with reciprocal rank 56.25\%, y7r7o3. 

\par \textbf{Level Skipping}: To test the effect of skipping levels, even levels (i.e., 2, 4, etc.) were omitted.
%to test whether  subsequent levels contain similar information. 
Odd levels were not skipped, in part because this matches the baseline X1 (bag of words) model. Configurations where only the final level for each region type was kept (i.e., the largest number of regions for each type) were also examined to observe how much information is provided in the final levels alone. %X1 (BoW), even skipping and only using the last level give us conditions which incrementally work towards answering our research question. 
By comparing all configurations with the bag of words, we can see whether adding spatial information yields significant differences, and by comparing configurations with omitted levels to the original configurations, we can observe whether subsequent levels contain redundant information, and how the finest-grained region partitioning perform on their own.
%to any significant differences as well as whether or not we need any levels other than the final ones. 
%\rz{The experiment does not consider individual levels -- this claim seems incorrect.}

%rjb specs: Ubuntu 20.04 machine with a 64 core Intel(R) Xeon(R) Gold 6326 CPU @ 2.90GHz, an NVIDIA A40 GPU and 3.5TB of disk space.

 \par {\bf Implementation.} 
 We take TREC run files produced by running our experiments, and load them into PyTerrier\footnote{https://github.com/terrier-org/pyterrier} to perform our statistical analysis.
PyTerrier also reports queries that perform better or worse relative to the baseline in each condition. We ran the experiments on a Ubuntu 20.04 machine with a 64 core Intel(R) Xeon(R) Gold 6326 CPU @ 2.90GHz, an NVIDIA A40 GPU and 3.5TB of disk space. An OpenSearch instance was instantiated on this system using 10 shards.

\par \textbf{Results}. 
%Results for our comparison of the single region (X1) bag of words (BoW) model to full PHOC configurations can be seen in Table \ref{tab:bow}.
% REMOVE to keep focus.
%while the comparison of full PHOC configurations to configurations with omitted levels can be seen in Table \ref{tab:skipping}. 
%Table \ref{tab:times} shows the associated index sizes on disk along with mean query retrieval times. 
All PHOC configurations are used to index the full ARQMath formula test collection, 
and 100 test queries were run from the 2022 ARQMath Task 2 topic file. We used Bonferroni-corrected T-tests with a required significance level of $p=0.05$ for each experiment. 
%Bonferroni makes a correction for increased error likelihood in detecting statistical differences when comparing multiple pairs of conditions. 
%All stats tests consider the distribution of metric values, including any improvements/weakening for individual queries, each of which produce a metric value in the distribution.
In Tables \ref{tab:bow} and \ref{tab:skipping} we use an asterisk (*) to denote metrics which are statistically significant from the baseline.

\par In Table \ref{tab:bow} we clearly see that all metrics are significantly higher than the BoW baseline for yr7, yr7o3 and x5y3r9. This provides strong evidence that adding spatial information to the BoW model is beneficial, as well as using rectangular regions. As seen in Table \ref{tab:times}, this additional effectiveness requires an increase in storage, as each configuration stores more than 50 bits than the BoW. For example, the BoW index takes up 985.2 MB on disk compared to 3.5 GB for x5y3r9. While each of these configurations is significantly different than the BoW baseline, there is not as much difference amongst them. When using the next smallest model in terms of bits, yr7, as a baseline and testing for statistical significance we find that only x5y3r9 sees a significant decrease in metrics, and only for $nDCG'$ and $MAP'$.
\par In Table \ref{tab:skipping}, we see that for each full PHOC configuration, when compared with their corresponding odd-levels-only and last-level-only counterparts, omitting levels impacts metrics measured on full rankings, but is less likely to affect shallow precision-based metrics. yr7o3-last decreased by $\sim2.6$ percent for $P'@10$ relative to yr7o3-full, which is notable as it was the only metric to see a significant change in regards to precision-based metrics. Even with this decrease for $P'@10$, yr7o3-last obtains the highest $P'@1$ out of all configurations, showing that even though there was a significant drop for $P'@10$, omitting all but the last level had little effect on precision overall. 
% RZ: UNCLEAR
%The results related before we start to see a consistent degradation in shallow metrics. 
The final levels of each configuration will have the largest number and smallest region sizes, and thus will reward exact matches, which will tend to increase precision because exact matches are usually highly relevant.

\par When analysing metrics at lower ranks ($nDCG'@1000$ and $MAP'$) we see multiple configurations where there is a significant decrease. yr7-odd, yr7-last, yr7o3-odd and yr7o3-last all produce lower $nDCG'@1000$ and $MAP'$ scores, while x5y3r9-odd shows a decrease in $nDCG'@1000$. Compared to the shallow metrics we see a more consistent drop in metrics, especially with configurations which only keep the last level. This likely means that these early levels are important for collecting formulas which are partially relevant and may appear further down the ranking list. This makes sense: if a formula shares certain symbols with a query (e.g., for a specific fraction), but are in a different spatial locations, the lower levels will provide additional matches in their larger regions.

\par Table \ref{tab:times} shows
%the mean retrieval time (MRT) and index size for all indices. As 
how index sizes on disk increase with the number of bits in a PHOC model.
%which is expected as these extra bits need to be stored somewhere. 
Retrieval times (MRT) are consistent across configurations,  because in our implementation each symbol PHOC vector is stored and used for scoring in a 64 bit long integer within OpenSearch. 
% The slight reduction might be skippable.
%\rz{The slight reduction in retrieval times for X1 (bag of words) is likely due to a reduction in the amount of data communicated between XXX and YYY.}

\begin{table}
    \centering
  \caption{Comparing Bag of Words (X1) and different PHOC configurations. * indicates %Metrics with a "*" next to them represent a 
  statistically significant results.}
  \label{tab:bow}
  \resizebox{\columnwidth}{!}{
  \begin{tabular}{l l c | c c  | c c c c c}
    \toprule
    \textbf{Model} & \textbf{Levels} & \textbf{Bits} & \textbf{NDCG$'$} & \textbf{MAP$'$} & \textbf{P$'$@10} & \textbf{P$'$@5} & \textbf{P$'$@1}\\
    \midrule
    \underline{X1 (BoW)} & Full & 1 & 0.4651 & 0.2947 & 0.5171 & 0.5605 & 0.6316\\
    yr7 & Full & 55 & 0.6228* & 0.4336* & 0.6092* & 0.6605* & 0.8026*\\    
    yr7o3 & Full & 60 & 0.6227* & 0.4338* & 0.6092* & 0.6711* & 0.7895*\\    
    x5y3r9 & Full & 64 & 0.5904* & 0.4024* & 0.5974* & 0.6816* & 0.8026*\\    
    \bottomrule
    \end{tabular}}
\end{table}

\begin{table}
    \centering
  \caption{Effects Of Level Skipping. Underlined models represent baselines. 
  * indicates
  %Metrics with a "*" next to them represent a 
  statistically significant results.}
  \label{tab:skipping}
  \resizebox{\columnwidth}{!}{
  \begin{tabular}{l l c | c c | c c c c c}
    \toprule    
    \textbf{Model} & \textbf{Levels} & \textbf{Bits} & \textbf{NDCG$'$} & \textbf{MAP$'$} & \textbf{P$'$@10} & \textbf{P$'$@5} & \textbf{P$'$@1}\\
    \midrule    
    \underline{yr7} & Full & 55 & 0.6228 & 0.4336 & 0.6092 & 0.6605 & 0.8026\\
    yr7 & Odd & 31 & 0.6183 & 0.4282 & 0.6092 & 0.6632 & 0.7500\\
    yr7 & Last & 14 & 0.5792* & 0.3914* & 0.5868 & 0.6447 & 0.7763\\
    \midrule
    \underline{yr7o3} & Full & 60 & 0.6227 & 0.4338 & 0.6092 & 0.6711 & 0.7895\\
    yr7o3 & Odd & 34 & 0.6180 & 0.4280 & 0.6053 & 0.6711 & 0.7763\\
    yr7o3 & Last & 17 & 0.5837* & 0.3945* & 0.5829* & 0.6526 & 0.8158\\
    \midrule
    \underline{x5y3r9} & Full & 64 & 0.5904 & 0.4024 & 0.5974 & 0.6816 & 0.8026\\
    x5y3r9 & Odd & 36 & 0.5833* & 0.3989 & 0.5947 & 0.6711 & 0.8026\\
    x5y3r9 & Last & 17 & 0.5813 & 0.3974 & 0.5961 & 0.6684 & 0.8026\\
    \bottomrule
    \end{tabular}
    }
\end{table}

\begin{table}
    \centering
  \caption{Index sizes and retrieval times for all models.}
  \label{tab:times}
  
  \begin{tabular}{l l l | c c }
    \toprule    
    \textbf{Model} & \textbf{Levels} & \textbf{Bits} & \textbf{Index Size (GB)} & \textbf{MRT (Seconds)} \\
    \midrule
    \underline{X1 (BoW)} & Full & 1 & 0.99 & 0.70\\
    \midrule    
    \underline{yr7} & Full & 55 &  2.6 & 0.76 \\
    yr7 & Odd & 31 & 2.1 & 0.77\\
    yr7 & Last & 14 & 1.6 & 0.75\\
    \midrule
    \underline{yr7o3} & Full & 60  & 3 & 0.76\\
    yr7o3 & Odd & 34 & 2.3 & 0.75\\
    yr7o3 & Last & 17 & 1.7 & 0.75 \\
    \midrule
    \underline{x5y3r9} & Full & 64 & 3.5 & 0.74\\
    x5y3r9 & Odd & 36 & 2.4 & 0.72\\
    x5y3r9 & Last & 17 & 1.8 & 0.74\\
    \bottomrule
    \end{tabular}
\end{table}

%\rz{MISSING: Retrieval speeds; Bits are shown in the table, but this is not mentioned, even though the 'Odd' models are not generally significantly different than the 'full' models (but double-check P'@1), and often the 'last' models aren't either. Do the smaller size models retrieve faster? }

%\rz{MISSING: Comparison with benchmark systems for ARQMath-3 (Robin and other previous PHOC models in particular). Many reviewers will want this in a separate table, e.g., comparing X1, one or more of the 'best' configurations, and the 'state-of-the-art' (i.e., best performing systems on ARQMath-3). }
\par The best performing system for Task 2 in ARQMath-3 was Approach0 \cite{approach02022}; having scores of $72\%$ for $nDCG'$, $56.8\%$ for $MAP'$ and $68.8\%$ for $P'@10$. The best PHOC-based model presented in this competition (xy7o4) scored $47.2\%$ for $nDCG'$, $30.9\%$ for $MAP'$ and $56.8\%$ for $P'@10$. Our smallest full PHOC configuration tested (yr7) produces improvements of over 10\% for $nDCG'$ and $MAP'$, and roughly 4\% for  $P'@10$.
%improved PHOC-based models performance by scoring $62.28\%$ for $nDCG'$, $43.36\%$ for $MAP'$, and $60.92\%$ for $P'@10$. 
While xy7 has weaker effectiveness  compared to Approach0 (reductions of $9.72\%$, $13.44\%$, and $7.88\%$ for $nDCG'$, $MAP'$ and $P'@10$ respectively), these differences are smaller than expected given the simplicity of the PHOC model, which matches specific symbols in space without synonymns (e.g., variable substitution) or weighting (e.g., symbol IDF). In contrast, A0 combines dense and sparse techniques in its retrieval model.

\section{Conclusion}
 We introduce rectangular regions for PHOC math formula retrieval and analyze the effect of omitting levels in  PHOC models. 
 %Comparing three configurations selected using a preliminary grid search with a Bag of Words model, 
 We demonstrate that rectangular regions help improve earlier PHOC models, and that the resulting models obtain 
 strong results relative to state-of-the-art formula retrieval models in the ARQMath-3 formula retrieval task. We also
 show that 
 %their additional spatial brings significant gains in retrieval effectiveness, improvements over previous PHOC models, and Further, we show that 
 while additional spatial information from PHOCs with larger numbers of levels is helpful, it is unnecessary to encode every level, as similar results can be obtained using fewer spatial regions. Future work includes applying PHOC models in neural embeddings and retrieval models.

\begin{acks}
This work was partially supported by the Alfred P. Sloan Foundation under Grant No. G-2017-9827 and the National Science Foundation (USA) under Grant No. IIS-1717997.
\end{acks}

\bibliographystyle{ACM-Reference-Format}
\bibliography{sample-base}


\begin{thebibliography}{31}

%%% the Web address.

\ifx \showCODEN    \undefined \def \showCODEN     #1{\unskip}     \fi
\ifx \showDOI      \undefined \def \showDOI       #1{#1}\fi
\ifx \showISBNx    \undefined \def \showISBNx     #1{\unskip}     \fi
\ifx \showISBNxiii \undefined \def \showISBNxiii  #1{\unskip}     \fi
\ifx \showISSN     \undefined \def \showISSN      #1{\unskip}     \fi
\ifx \showLCCN     \undefined \def \showLCCN      #1{\unskip}     \fi
\ifx \shownote     \undefined \def \shownote      #1{#1}          \fi
\ifx \showarticletitle \undefined \def \showarticletitle #1{#1}   \fi
\ifx \showURL      \undefined \def \showURL       {\relax}        \fi
% invisible to TeX
\providecommand\bibfield[2]{#2}
\providecommand\bibinfo[2]{#2}
\providecommand\natexlab[1]{#1}
\providecommand\showeprint[2][]{arXiv:#2}

\bibitem[Almaz{\'{a}}n et~al\mbox{.}(2013)]%
        {zero_phoc}
\bibfield{author}{\bibinfo{person}{Jon Almaz{\'{a}}n}, \bibinfo{person}{Albert
  Gordo}, \bibinfo{person}{Alicia Forn{\'{e}}s}, {and} \bibinfo{person}{Ernest
  Valveny}.} \bibinfo{year}{2013}\natexlab{}.
\newblock \showarticletitle{Handwritten Word Spotting with Corrected
  Attributes}. In \bibinfo{booktitle}{\emph{{ICCV}}}.
  \bibinfo{publisher}{{IEEE} Computer Society}, \bibinfo{pages}{1017--1024}.
\newblock


\bibitem[Almaz{\'{a}}n et~al\mbox{.}(2014)]%
        {first_phoc}
\bibfield{author}{\bibinfo{person}{Jon Almaz{\'{a}}n}, \bibinfo{person}{Albert
  Gordo}, \bibinfo{person}{Alicia Forn{\'{e}}s}, {and} \bibinfo{person}{Ernest
  Valveny}.} \bibinfo{year}{2014}\natexlab{}.
\newblock \showarticletitle{Word Spotting and Recognition with Embedded
  Attributes}.
\newblock \bibinfo{journal}{\emph{{IEEE} Trans. Pattern Anal. Mach. Intell.}}
  \bibinfo{volume}{36}, \bibinfo{number}{12} (\bibinfo{year}{2014}),
  \bibinfo{pages}{2552--2566}.
\newblock
\urldef\tempurl%
\url{https://doi.org/10.1109/TPAMI.2014.2339814}
\showDOI{\tempurl}


\bibitem[Avenoso(2021)]%
        {avenoso_spatial_2021}
\bibfield{author}{\bibinfo{person}{Robin Avenoso}.}
  \bibinfo{year}{2021}\natexlab{}.
\newblock \showarticletitle{Spatial vs. {Graph}-{Based} {Formula} {Retrieval}}.
\newblock \bibinfo{journal}{\emph{Theses}} (\bibinfo{date}{May}
  \bibinfo{year}{2021}).
\newblock
\urldef\tempurl%
\url{https://scholarworks.rit.edu/theses/10784}
\showURL{%
\tempurl}


\bibitem[Avenoso et~al\mbox{.}(2021)]%
        {avenoso_xy-phoc_2021}
\bibfield{author}{\bibinfo{person}{Robin Avenoso}, \bibinfo{person}{Behrooz
  Mansouri}, {and} \bibinfo{person}{Richard Zanibbi}.}
  \bibinfo{year}{2021}\natexlab{}.
\newblock \showarticletitle{{XY}-{PHOC} {Symbol} {Location} {Embeddings} for
  {Math} {Formula} {Retrieval} and {Autocompletion}}. In
  \bibinfo{booktitle}{\emph{Proceedings of the {Working} {Notes} of {CLEF} 2021
  - {Conference} and {Labs} of the {Evaluation} {Forum}}}
  \emph{(\bibinfo{series}{{CEUR} {Workshop} {Proceedings}},
  Vol.~\bibinfo{volume}{2936})}, \bibfield{editor}{\bibinfo{person}{Guglielmo
  Faggioli}, \bibinfo{person}{Nicola Ferro}, \bibinfo{person}{Alexis Joly},
  \bibinfo{person}{Maria Maistro}, {and} \bibinfo{person}{Florina Piroi}}
  (Eds.). \bibinfo{publisher}{CEUR}, \bibinfo{address}{Bucharest, Romania},
  \bibinfo{pages}{25--35}.
\newblock
\urldef\tempurl%
\url{http://ceur-ws.org/Vol-2936/#paper-02}
\showURL{%
\tempurl}
\newblock
\shownote{ISSN: 1613-0073}.


\bibitem[Bender et~al\mbox{.}(2021)]%
        {parrot}
\bibfield{author}{\bibinfo{person}{Emily~M. Bender}, \bibinfo{person}{Timnit
  Gebru}, \bibinfo{person}{Angelina McMillan{-}Major}, {and}
  \bibinfo{person}{Shmargaret Shmitchell}.} \bibinfo{year}{2021}\natexlab{}.
\newblock \showarticletitle{On the Dangers of Stochastic Parrots: Can Language
  Models Be Too Big?}. In \bibinfo{booktitle}{\emph{FAccT '21: 2021 {ACM}
  Conference on Fairness, Accountability, and Transparency, Virtual Event /
  Toronto, Canada, March 3-10, 2021}},
  \bibfield{editor}{\bibinfo{person}{Madeleine~Clare Elish},
  \bibinfo{person}{William Isaac}, {and} \bibinfo{person}{Richard~S. Zemel}}
  (Eds.). \bibinfo{publisher}{{ACM}}, \bibinfo{pages}{610--623}.
\newblock
\urldef\tempurl%
\url{https://doi.org/10.1145/3442188.3445922}
\showDOI{\tempurl}


\bibitem[Castellanos(2017)]%
        {kennythesis}
\bibfield{author}{\bibinfo{person}{Kenny~Davila Castellanos}.}
  \bibinfo{year}{2017}\natexlab{}.
\newblock \showarticletitle{Symbolic and Visual Retrieval of Mathematical
  Notation using Formula Graph Symbol Pair Matching and Structural Alignment}.
\newblock


\bibitem[Davila and Zanibbi(2017)]%
        {tangentS}
\bibfield{author}{\bibinfo{person}{Kenny Davila} {and} \bibinfo{person}{Richard
  Zanibbi}.} \bibinfo{year}{2017}\natexlab{}.
\newblock \showarticletitle{Layout and Semantics: Combining Representations for
  Mathematical Formula Search}. In \bibinfo{booktitle}{\emph{Proceedings of the
  40th International {ACM} {SIGIR} Conference on Research and Development in
  Information Retrieval, Shinjuku, Tokyo, Japan, August 7-11, 2017}},
  \bibfield{editor}{\bibinfo{person}{Noriko Kando}, \bibinfo{person}{Tetsuya
  Sakai}, \bibinfo{person}{Hideo Joho}, \bibinfo{person}{Hang Li},
  \bibinfo{person}{Arjen~P. de~Vries}, {and} \bibinfo{person}{Ryen~W. White}}
  (Eds.). \bibinfo{publisher}{{ACM}}, \bibinfo{pages}{1165--1168}.
\newblock
\urldef\tempurl%
\url{https://doi.org/10.1145/3077136.3080748}
\showDOI{\tempurl}


\bibitem[Fraser et~al\mbox{.}(2018)]%
        {tompaBM25}
\bibfield{author}{\bibinfo{person}{Dallas~J. Fraser}, \bibinfo{person}{Andrew
  Kane}, {and} \bibinfo{person}{Frank~Wm. Tompa}.}
  \bibinfo{year}{2018}\natexlab{}.
\newblock \showarticletitle{Choosing Math Features for {BM25} Ranking with
  Tangent-L}. In \bibinfo{booktitle}{\emph{Proceedings of the {ACM} Symposium
  on Document Engineering 2018, DocEng 2018, Halifax, NS, Canada, August 28-31,
  2018}}. \bibinfo{publisher}{{ACM}}, \bibinfo{pages}{17:1--17:10}.
\newblock
\urldef\tempurl%
\url{https://doi.org/10.1145/3209280.3209527}
\showDOI{\tempurl}


\bibitem[Gao et~al\mbox{.}(2017)]%
        {symbol2vec}
\bibfield{author}{\bibinfo{person}{Liangcai Gao}, \bibinfo{person}{Zhuoren
  Jiang}, \bibinfo{person}{Yue Yin}, \bibinfo{person}{Ke Yuan},
  \bibinfo{person}{Zuoyu Yan}, {and} \bibinfo{person}{Zhi Tang}.}
  \bibinfo{year}{2017}\natexlab{}.
\newblock \showarticletitle{Preliminary Exploration of Formula Embedding for
  Mathematical Information Retrieval: can mathematical formulae be embedded
  like a natural language?}
\newblock \bibinfo{journal}{\emph{CoRR}}  \bibinfo{volume}{abs/1707.05154}
  (\bibinfo{year}{2017}).
\newblock
\showeprint[arXiv]{1707.05154}
\urldef\tempurl%
\url{http://arxiv.org/abs/1707.05154}
\showURL{%
\tempurl}


\bibitem[Ghosh and Valveny(2015)]%
        {second_phoc}
\bibfield{author}{\bibinfo{person}{Suman~K. Ghosh} {and}
  \bibinfo{person}{Ernest Valveny}.} \bibinfo{year}{2015}\natexlab{}.
\newblock \showarticletitle{Query by string word spotting based on character
  bi-gram indexing}. In \bibinfo{booktitle}{\emph{{ICDAR}}}.
  \bibinfo{publisher}{{IEEE} Computer Society}, \bibinfo{pages}{881--885}.
\newblock


\bibitem[Guidi and Coen(2016)]%
        {survey1}
\bibfield{author}{\bibinfo{person}{Ferruccio Guidi} {and}
  \bibinfo{person}{Claudio~Sacerdoti Coen}.} \bibinfo{year}{2016}\natexlab{}.
\newblock \showarticletitle{A Survey on Retrieval of Mathematical Knowledge}.
\newblock \bibinfo{journal}{\emph{Math. Comput. Sci.}} \bibinfo{volume}{10},
  \bibinfo{number}{4} (\bibinfo{year}{2016}), \bibinfo{pages}{409--427}.
\newblock
\urldef\tempurl%
\url{https://doi.org/10.1007/s11786-016-0274-0}
\showDOI{\tempurl}


\bibitem[Kristianto et~al\mbox{.}(2016)]%
        {kristianto2016mcat}
\bibfield{author}{\bibinfo{person}{Giovanni~Yoko Kristianto},
  \bibinfo{person}{Goran Topic}, {and} \bibinfo{person}{Akiko Aizawa}.}
  \bibinfo{year}{2016}\natexlab{}.
\newblock \showarticletitle{MCAT Math Retrieval System for NTCIR-12 MathIR
  Task.}. In \bibinfo{booktitle}{\emph{NTCIR}}.
\newblock


\bibitem[Langsenkamp et~al\mbox{.}(2022)]%
        {DBLP:conf/clef/LangsenkampMZ22}
\bibfield{author}{\bibinfo{person}{Matt Langsenkamp}, \bibinfo{person}{Behrooz
  Mansouri}, {and} \bibinfo{person}{Richard Zanibbi}.}
  \bibinfo{year}{2022}\natexlab{}.
\newblock \showarticletitle{Expanding Spatial Regions and Incorporating {IDF}
  for PHOC-Based Math Formula Retrieval at ARQMath-3}. In
  \bibinfo{booktitle}{\emph{Proceedings of the Working Notes of {CLEF} 2022 -
  Conference and Labs of the Evaluation Forum, Bologna, Italy, September 5th -
  to - 8th, 2022}} \emph{(\bibinfo{series}{{CEUR} Workshop Proceedings},
  Vol.~\bibinfo{volume}{3180})}, \bibfield{editor}{\bibinfo{person}{Guglielmo
  Faggioli}, \bibinfo{person}{Nicola Ferro}, \bibinfo{person}{Allan Hanbury},
  {and} \bibinfo{person}{Martin Potthast}} (Eds.).
  \bibinfo{publisher}{CEUR-WS.org}, \bibinfo{pages}{63--82}.
\newblock
\urldef\tempurl%
\url{http://ceur-ws.org/Vol-3180/paper-04.pdf}
\showURL{%
\tempurl}


\bibitem[Mansouri et~al\mbox{.}(2022a)]%
        {ARQ3}
\bibfield{author}{\bibinfo{person}{Behrooz Mansouri},
  \bibinfo{person}{V{\'{\i}}t Novotn{\'{y}}}, \bibinfo{person}{Anurag Agarwal},
  \bibinfo{person}{Douglas~W. Oard}, {and} \bibinfo{person}{Richard Zanibbi}.}
  \bibinfo{year}{2022}\natexlab{a}.
\newblock \showarticletitle{Overview of ARQMath-3 {(2022):} Third {CLEF} Lab on
  Answer Retrieval for Questions on Math}. In
  \bibinfo{booktitle}{\emph{Experimental {IR} Meets Multilinguality,
  Multimodality, and Interaction - 13th International Conference of the {CLEF}
  Association, {CLEF} 2022, Bologna, Italy, September 5-8, 2022, Proceedings}}
  \emph{(\bibinfo{series}{Lecture Notes in Computer Science},
  Vol.~\bibinfo{volume}{13390})}, \bibfield{editor}{\bibinfo{person}{Alberto
  Barr{\'{o}}n{-}Cede{\~{n}}o}, \bibinfo{person}{Giovanni Da~San Martino},
  \bibinfo{person}{Mirko~Degli Esposti}, \bibinfo{person}{Fabrizio Sebastiani},
  \bibinfo{person}{Craig Macdonald}, \bibinfo{person}{Gabriella Pasi},
  \bibinfo{person}{Allan Hanbury}, \bibinfo{person}{Martin Potthast},
  \bibinfo{person}{Guglielmo Faggioli}, {and} \bibinfo{person}{Nicola Ferro}}
  (Eds.). \bibinfo{publisher}{Springer}, \bibinfo{pages}{286--310}.
\newblock
\urldef\tempurl%
\url{https://doi.org/10.1007/978-3-031-13643-6\_20}
\showDOI{\tempurl}


\bibitem[Mansouri et~al\mbox{.}(2022b)]%
        {DBLP:conf/clef/MansouriNAOZ22}
\bibfield{author}{\bibinfo{person}{Behrooz Mansouri},
  \bibinfo{person}{V{\'{\i}}t Novotn{\'{y}}}, \bibinfo{person}{Anurag Agarwal},
  \bibinfo{person}{Douglas~W. Oard}, {and} \bibinfo{person}{Richard Zanibbi}.}
  \bibinfo{year}{2022}\natexlab{b}.
\newblock \showarticletitle{Overview of ARQMath-3 {(2022):} Third {CLEF} Lab on
  Answer Retrieval for Questions on Math (Working Notes Version)}. In
  \bibinfo{booktitle}{\emph{Proceedings of the Working Notes of {CLEF} 2022 -
  Conference and Labs of the Evaluation Forum, Bologna, Italy, September 5th -
  to - 8th, 2022}} \emph{(\bibinfo{series}{{CEUR} Workshop Proceedings},
  Vol.~\bibinfo{volume}{3180})}, \bibfield{editor}{\bibinfo{person}{Guglielmo
  Faggioli}, \bibinfo{person}{Nicola Ferro}, \bibinfo{person}{Allan Hanbury},
  {and} \bibinfo{person}{Martin Potthast}} (Eds.).
  \bibinfo{publisher}{CEUR-WS.org}, \bibinfo{pages}{1--27}.
\newblock
\urldef\tempurl%
\url{http://ceur-ws.org/Vol-3180/paper-01.pdf}
\showURL{%
\tempurl}


\bibitem[Mansouri et~al\mbox{.}(2021a)]%
        {mathamr}
\bibfield{author}{\bibinfo{person}{Behrooz Mansouri},
  \bibinfo{person}{Douglas~W Oard}, {and} \bibinfo{person}{Richard Zanibbi}.}
  \bibinfo{year}{2021}\natexlab{a}.
\newblock \showarticletitle{DPRL Systems in the CLEF 2022 ARQMath Lab:
  Introducing MathAMR for Math-Aware Search}.
\newblock \bibinfo{journal}{\emph{Proceedings of the Working Notes of CLEF
  2022}} (\bibinfo{year}{2021}), \bibinfo{pages}{5--8}.
\newblock


\bibitem[Mansouri et~al\mbox{.}(2019)]%
        {tangentCFT}
\bibfield{author}{\bibinfo{person}{Behrooz Mansouri}, \bibinfo{person}{Shaurya
  Rohatgi}, \bibinfo{person}{Douglas~W. Oard}, \bibinfo{person}{Jian Wu},
  \bibinfo{person}{C.~Lee Giles}, {and} \bibinfo{person}{Richard Zanibbi}.}
  \bibinfo{year}{2019}\natexlab{}.
\newblock \showarticletitle{Tangent-CFT: An Embedding Model for Mathematical
  Formulas}. In \bibinfo{booktitle}{\emph{Proceedings of the 2019 {ACM} {SIGIR}
  International Conference on Theory of Information Retrieval, {ICTIR} 2019,
  Santa Clara, CA, USA, October 2-5, 2019}},
  \bibfield{editor}{\bibinfo{person}{Yi~Fang}, \bibinfo{person}{Yi~Zhang},
  \bibinfo{person}{James Allan}, \bibinfo{person}{Krisztian Balog},
  \bibinfo{person}{Ben Carterette}, {and} \bibinfo{person}{Jiafeng Guo}}
  (Eds.). \bibinfo{publisher}{{ACM}}, \bibinfo{pages}{11--18}.
\newblock
\urldef\tempurl%
\url{https://doi.org/10.1145/3341981.3344235}
\showDOI{\tempurl}


\bibitem[Mansouri et~al\mbox{.}(2021b)]%
        {ARQ2021}
\bibfield{author}{\bibinfo{person}{Behrooz Mansouri}, \bibinfo{person}{Richard
  Zanibbi}, \bibinfo{person}{Douglas~W. Oard}, {and} \bibinfo{person}{Anurag
  Agarwal}.} \bibinfo{year}{2021}\natexlab{b}.
\newblock \showarticletitle{Overview of ARQMath-2 {(2021):} Second {CLEF} Lab
  on Answer Retrieval for Questions on Math}. In
  \bibinfo{booktitle}{\emph{Experimental {IR} Meets Multilinguality,
  Multimodality, and Interaction - 12th International Conference of the {CLEF}
  Association, {CLEF} 2021, Virtual Event, September 21-24, 2021, Proceedings}}
  \emph{(\bibinfo{series}{Lecture Notes in Computer Science},
  Vol.~\bibinfo{volume}{12880})},
  \bibfield{editor}{\bibinfo{person}{K.~Sel{\c{c}}uk Candan},
  \bibinfo{person}{Bogdan Ionescu}, \bibinfo{person}{Lorraine Goeuriot},
  \bibinfo{person}{Birger Larsen}, \bibinfo{person}{Henning M{\"{u}}ller},
  \bibinfo{person}{Alexis Joly}, \bibinfo{person}{Maria Maistro},
  \bibinfo{person}{Florina Piroi}, \bibinfo{person}{Guglielmo Faggioli}, {and}
  \bibinfo{person}{Nicola Ferro}} (Eds.). \bibinfo{publisher}{Springer},
  \bibinfo{pages}{215--238}.
\newblock
\urldef\tempurl%
\url{https://doi.org/10.1007/978-3-030-85251-1\_17}
\showDOI{\tempurl}


\bibitem[Miller and Youssef(2003)]%
        {DBLP:journals/amai/MillerY03}
\bibfield{author}{\bibinfo{person}{Bruce~R. Miller} {and}
  \bibinfo{person}{Abdou Youssef}.} \bibinfo{year}{2003}\natexlab{}.
\newblock \showarticletitle{Technical Aspects of the Digital Library of
  Mathematical Functions}.
\newblock \bibinfo{journal}{\emph{Ann. Math. Artif. Intell.}}
  \bibinfo{volume}{38}, \bibinfo{number}{1-3} (\bibinfo{year}{2003}),
  \bibinfo{pages}{121--136}.
\newblock
\urldef\tempurl%
\url{https://doi.org/10.1023/A:1022967814992}
\showDOI{\tempurl}


\bibitem[Peng et~al\mbox{.}(2021)]%
        {mathbert}
\bibfield{author}{\bibinfo{person}{Shuai Peng}, \bibinfo{person}{Ke Yuan},
  \bibinfo{person}{Liangcai Gao}, {and} \bibinfo{person}{Zhi Tang}.}
  \bibinfo{year}{2021}\natexlab{}.
\newblock \showarticletitle{MathBERT: {A} Pre-Trained Model for Mathematical
  Formula Understanding}.
\newblock \bibinfo{journal}{\emph{CoRR}}  \bibinfo{volume}{abs/2105.00377}
  (\bibinfo{year}{2021}).
\newblock
\showeprint[arXiv]{2105.00377}
\urldef\tempurl%
\url{https://arxiv.org/abs/2105.00377}
\showURL{%
\tempurl}


\bibitem[Sakai and Kando(2008)]%
        {sakai}
\bibfield{author}{\bibinfo{person}{Tetsuya Sakai} {and} \bibinfo{person}{Noriko
  Kando}.} \bibinfo{year}{2008}\natexlab{}.
\newblock \showarticletitle{On information retrieval metrics designed for
  evaluation with incomplete relevance assessments}.
\newblock \bibinfo{journal}{\emph{Inf. Retr.}} \bibinfo{volume}{11},
  \bibinfo{number}{5} (\bibinfo{year}{2008}), \bibinfo{pages}{447--470}.
\newblock
\urldef\tempurl%
\url{https://doi.org/10.1007/s10791-008-9059-7}
\showDOI{\tempurl}


\bibitem[Shah et~al\mbox{.}(2021)]%
        {sscraper}
\bibfield{author}{\bibinfo{person}{Ayush~Kumar Shah}, \bibinfo{person}{Abhisek
  Dey}, {and} \bibinfo{person}{Richard Zanibbi}.}
  \bibinfo{year}{2021}\natexlab{}.
\newblock \showarticletitle{A {Math} {Formula} {Extraction} and {Evaluation}
  {Framework} for {PDF} {Documents}}.
\newblock In \bibinfo{booktitle}{\emph{Document {Analysis} and {Recognition}
  – {ICDAR} 2021}}, \bibfield{editor}{\bibinfo{person}{Josep Lladós},
  \bibinfo{person}{Daniel Lopresti}, {and} \bibinfo{person}{Seiichi Uchida}}
  (Eds.). Vol.~\bibinfo{volume}{12822}. \bibinfo{publisher}{Springer
  International Publishing}, \bibinfo{address}{Cham}, \bibinfo{pages}{19--34}.
\newblock
\showISBNx{978-3-030-86330-2 978-3-030-86331-9}
\urldef\tempurl%
\url{https://doi.org/10.1007/978-3-030-86331-9_2}
\showDOI{\tempurl}
\newblock
\shownote{Series Title: Lecture Notes in Computer Science}.


\bibitem[Sojka and L{\'{\i}}ska(2011)]%
        {art}
\bibfield{author}{\bibinfo{person}{Petr Sojka} {and} \bibinfo{person}{Martin
  L{\'{\i}}ska}.} \bibinfo{year}{2011}\natexlab{}.
\newblock \showarticletitle{The art of mathematics retrieval}. In
  \bibinfo{booktitle}{\emph{Proceedings of the 2011 {ACM} Symposium on Document
  Engineering, Mountain View, CA, USA, September 19-22, 2011}},
  \bibfield{editor}{\bibinfo{person}{Matthew R.~B. Hardy} {and}
  \bibinfo{person}{Frank~Wm. Tompa}} (Eds.). \bibinfo{publisher}{{ACM}},
  \bibinfo{pages}{57--60}.
\newblock
\urldef\tempurl%
\url{https://doi.org/10.1145/2034691.2034703}
\showDOI{\tempurl}


\bibitem[Stalnaker and Zanibbi(2015)]%
        {Tangent}
\bibfield{author}{\bibinfo{person}{David Stalnaker} {and}
  \bibinfo{person}{Richard Zanibbi}.} \bibinfo{year}{2015}\natexlab{}.
\newblock \showarticletitle{Math expression retrieval using an inverted index
  over symbol pairs}, \bibfield{editor}{\bibinfo{person}{Eric~K. Ringger} {and}
  \bibinfo{person}{Bart Lamiroy}} (Eds.). \bibinfo{address}{San Francisco,
  California, USA}, \bibinfo{pages}{940207}.
\newblock
\urldef\tempurl%
\url{https://doi.org/10.1117/12.2074084}
\showDOI{\tempurl}


\bibitem[Zanibbi and Blostein(2012)]%
        {survey2}
\bibfield{author}{\bibinfo{person}{Richard Zanibbi} {and}
  \bibinfo{person}{Dorothea Blostein}.} \bibinfo{year}{2012}\natexlab{}.
\newblock \showarticletitle{Recognition and retrieval of mathematical
  expressions}.
\newblock \bibinfo{journal}{\emph{Int. J. Document Anal. Recognit.}}
  \bibinfo{volume}{15}, \bibinfo{number}{4} (\bibinfo{year}{2012}),
  \bibinfo{pages}{331--357}.
\newblock
\urldef\tempurl%
\url{https://doi.org/10.1007/s10032-011-0174-4}
\showDOI{\tempurl}


\bibitem[Zanibbi et~al\mbox{.}(2001)]%
        {bst}
\bibfield{author}{\bibinfo{person}{Richard Zanibbi}, \bibinfo{person}{Dorothea
  Blostein}, {and} \bibinfo{person}{James~R. Cordy}.}
  \bibinfo{year}{2001}\natexlab{}.
\newblock \showarticletitle{Baseline structure analysis of handwritten
  mathematics notation}.
\newblock \bibinfo{journal}{\emph{Proceedings of Sixth International Conference
  on Document Analysis and Recognition}} (\bibinfo{year}{2001}),
  \bibinfo{pages}{768--773}.
\newblock


\bibitem[Zanibbi et~al\mbox{.}(2020)]%
        {arq2020}
\bibfield{author}{\bibinfo{person}{Richard Zanibbi},
  \bibinfo{person}{Douglas~W. Oard}, \bibinfo{person}{Anurag Agarwal}, {and}
  \bibinfo{person}{Behrooz Mansouri}.} \bibinfo{year}{2020}\natexlab{}.
\newblock \showarticletitle{Overview of ARQMath 2020 (Updated Working Notes
  Version): {CLEF} Lab on Answer Retrieval for Questions on Math}. In
  \bibinfo{booktitle}{\emph{Working Notes of {CLEF} 2020 - Conference and Labs
  of the Evaluation Forum, Thessaloniki, Greece, September 22-25, 2020}}
  \emph{(\bibinfo{series}{{CEUR} Workshop Proceedings},
  Vol.~\bibinfo{volume}{2696})}, \bibfield{editor}{\bibinfo{person}{Linda
  Cappellato}, \bibinfo{person}{Carsten Eickhoff}, \bibinfo{person}{Nicola
  Ferro}, {and} \bibinfo{person}{Aur{\'{e}}lie N{\'{e}}v{\'{e}}ol}} (Eds.).
  \bibinfo{publisher}{CEUR-WS.org}.
\newblock
\urldef\tempurl%
\url{http://ceur-ws.org/Vol-2696/paper\_271.pdf}
\showURL{%
\tempurl}


\bibitem[Zanibbi and Yuan(2011)]%
        {DBLP:conf/drr/ZanibbiY11}
\bibfield{author}{\bibinfo{person}{Richard Zanibbi} {and} \bibinfo{person}{Bo
  Yuan}.} \bibinfo{year}{2011}\natexlab{}.
\newblock \showarticletitle{Keyword and image-based retrieval of mathematical
  expressions}. In \bibinfo{booktitle}{\emph{Document Recognition and Retrieval
  XVIII, part of the IS{\&}T-SPIE Electronic Imaging Symposium, San Jose, CA,
  USA, January 26-27, 2011, Proceedings}} \emph{(\bibinfo{series}{{SPIE}
  Proceedings}, Vol.~\bibinfo{volume}{7874})},
  \bibfield{editor}{\bibinfo{person}{Gady Agam} {and}
  \bibinfo{person}{Christian Viard{-}Gaudin}} (Eds.).
  \bibinfo{publisher}{{SPIE}}, \bibinfo{pages}{78740I}.
\newblock
\urldef\tempurl%
\url{https://doi.org/10.1117/12.873312}
\showDOI{\tempurl}


\bibitem[Zhong et~al\mbox{.}(2022a)]%
        {colbertWEI}
\bibfield{author}{\bibinfo{person}{Wei Zhong}, \bibinfo{person}{Yuqing Xie},
  {and} \bibinfo{person}{Jimmy Lin}.} \bibinfo{year}{2022}\natexlab{a}.
\newblock \showarticletitle{Applying Structural and Dense Semantic Matching for
  the {ARQMath} Lab 2022, {CLEF}}. In \bibinfo{booktitle}{\emph{Proceedings of
  the Working Notes of {CLEF} 2022 - Conference and Labs of the Evaluation
  Forum, Bologna, Italy, September 5th - to - 8th, 2022}}
  \emph{(\bibinfo{series}{{CEUR} Workshop Proceedings},
  Vol.~\bibinfo{volume}{3180})}, \bibfield{editor}{\bibinfo{person}{Guglielmo
  Faggioli}, \bibinfo{person}{Nicola Ferro}, \bibinfo{person}{Allan Hanbury},
  {and} \bibinfo{person}{Martin Potthast}} (Eds.).
  \bibinfo{publisher}{CEUR-WS.org}, \bibinfo{pages}{147--170}.
\newblock
\urldef\tempurl%
\url{https://ceur-ws.org/Vol-3180/paper-09.pdf}
\showURL{%
\tempurl}


\bibitem[Zhong et~al\mbox{.}(2022b)]%
        {approach02022}
\bibfield{author}{\bibinfo{person}{Wei Zhong}, \bibinfo{person}{Yuqing Xie},
  {and} \bibinfo{person}{Jimmy Lin}.} \bibinfo{year}{2022}\natexlab{b}.
\newblock \showarticletitle{Applying Structural and Dense Semantic Matching for
  the ARQMath Lab 2022, CLEF}.
\newblock \bibinfo{journal}{\emph{Proceedings of the Working Notes of CLEF
  2022}} (\bibinfo{year}{2022}), \bibinfo{pages}{5--8}.
\newblock


\bibitem[Zhong et~al\mbox{.}(2021)]%
        {approach0}
\bibfield{author}{\bibinfo{person}{Wei Zhong}, \bibinfo{person}{Xinyu Zhang},
  \bibinfo{person}{Ji Xin}, \bibinfo{person}{Richard Zanibbi}, {and}
  \bibinfo{person}{Jimmy Lin}.} \bibinfo{year}{2021}\natexlab{}.
\newblock \showarticletitle{Approach Zero and Anserini at the {CLEF-2021}
  ARQMath Track: Applying Substructure Search and {BM25} on Operator Tree Path
  Tokens}. In \bibinfo{booktitle}{\emph{Proceedings of the Working Notes of
  {CLEF} 2021 - Conference and Labs of the Evaluation Forum, Bucharest,
  Romania, September 21st - to - 24th, 2021}} \emph{(\bibinfo{series}{{CEUR}
  Workshop Proceedings}, Vol.~\bibinfo{volume}{2936})},
  \bibfield{editor}{\bibinfo{person}{Guglielmo Faggioli},
  \bibinfo{person}{Nicola Ferro}, \bibinfo{person}{Alexis Joly},
  \bibinfo{person}{Maria Maistro}, {and} \bibinfo{person}{Florina Piroi}}
  (Eds.). \bibinfo{publisher}{CEUR-WS.org}, \bibinfo{pages}{133--156}.
\newblock
\urldef\tempurl%
\url{http://ceur-ws.org/Vol-2936/paper-09.pdf}
\showURL{%
\tempurl}


\end{thebibliography}

\appendix

\end{document}